**CNN-based Automatic Detection of Bone Conditions via Diagnostic CT Images for Osteoporosis Screening**

Chao Tang [1], Wenkun Zhang [1], Haiting Li [2], Lei Li [1], Ziheng Li [1], Ailong Cai [1], Linyuan Wang [1], Dapeng Shi [2], and Bin Yan [1]

[1] PLA Strategy Support Force Information Engineering University, Henan Province, China.

[2] Department of Radiology in Henan Provincial People's Hospital, Henan Province, China

**Corresponding author:** Bin Yan, PhD, Professor, PLA Strategy Support Force Information Engineering University, No.62 Science Avenue, Zhengzhou, Henan Province, China.

**Tel**: +86-18290294068. **E-mail**: ybspace@hotmail.com

**Acknowledgments:** We are grateful to all of the participants in this study as well as all of the technicians who performed the scans.

**Funding:** This work was supported by the Nation Key Research and Development Program of China under grant 2018YFC0114500.

## Abstract

**Purpose:** The purpose of this study is to design a novelty automatic diagnostic method for osteoporosis screening by using the potential capability of convolutional neural network (CNN) in feature representation and extraction, which can be incorporated into the procedure of routine CT diagnostic in physical examination thereby improving the osteoporosis diagnosis and reducing the patient burden.

**Methods:** The proposed convolutional neural network-based method mainly comprises two functional modules to perform automatic detection of bone condition by analyzing the diagnostic CT image. The first functional module aims to locate and segment the ROI of diagnostic CT image, called Mark-Segmentation-Network (MS-Net). The second functional module is used to determine the category of bone condition by the features of ROI, called Bone-Conditions-Classification-Network (BCC-Net). The trained MS-Net can get the mark image of inputted original CT image, thereby obtain the segmentation image. The trained BCC-Net can obtain the probability value of normal bone mass, low bone mass, and osteoporosis by inputting the segmentation image. On the basis of network results, the radiologists can provide preliminary qualitative diagnosis results of bone condition.

**Results:** The proposed MS-Net has an excellent segmentation precision on the shape preservation of different lumbar vertebra. The numerical indicators of segmentation results are greater than 0.998. The proposed BCC-Net was evaluated with 3,024 clinical images and achieved an accuracy of 76.65% and an area under the receiver operating characteristic curve of 0.9167.

**Conclusions:** This study proposed a novel two-step method for automatic detection of bone conditions via diagnostic CT images and it has great potential in clinical applications for osteoporosis screening. The method can potentially reduce the manual burden to radiologists and diagnostic cost to patients.

# 1. Introduction

Osteoporosis is a skeletal disease characterized by low bone mass and microarchitectural deterioration of bone tissue [1][2]. It is an important public health issue because of its potentially devastating results [3] and high cumulative rate of fractures [4]. From the perspective of patients, fractures and subsequent loss of autonomy often represent a major drop in quality of life. Additionally, osteoporotic fractures of the vertebra carry a 12-month excess mortality of up to 20% because they require hospitalization and have consequently enhanced risk of other complications, such as pneumonia or thromboembolic disease, due to chronic immobilization [5]. Furthermore, the loss of bone mass occurs insidiously and is initially asymptomatic; thus, osteoporosis is usually diagnosed after bone fracture [6][7], which results in substantial personal suffering and socio-economic burden. Therefore, osteoporosis must be detected and treated early to avoid fragility fractures.

For the diagnosis of osteoporosis, bone mineral density (BMD) is an important indicator that is used by the World Health Organization to define the diagnostic criteria for osteoporosis [8]. BMD measurement can provide a basis for the prevention of osteoporotic fractures [9][10][11]. At present, the main clinical methods for BMD measurement include dual-energy X-ray absorptiometry (DXA) [12] and quantitative computed tomography (QCT) [13]. DXA is the gold standard for BMD measurement [14]. The measurement result of BMD on DXA is a T score. On the basis of the T score magnitude, the bone condition can be determined [15]. For example, osteoporosis is determined by a T score of less than -2.5. The measurement result of BMD on QCT is volumetric BMD (vBMD), and the diagnosis of bone condition is based on the vBMD magnitude [16]. However, these methods require auxiliary hardware or workflow and will incur extra costs to the physical examinees; consequently, people are reluctant to measure their BMD frequently for detection of bone condition. Additionally, osteoporosis is often under-diagnosed and under-reported in CT examinations. Early screening for osteoporosis is difficult in clinical practice. One possible strategy to overcome these problems and improve the clinical diagnosis of osteoporosis is to develop an automatic detection method for bone condition using diagnostic CT images obtained for other reasons, especially during physical examinations. A preliminary qualitative diagnosis result of bone condition can be obtained by exploiting the feature information contained in routine diagnostic CT images.

In recent years, deep learning techniques have been developed to automatically process images for handling detection and classification tasks. Convolutional neural networks (CNNs) have shown excellent performance in image segmentation [17][18][19][20] and image classification [21][22][23][24] because of its powerful feature representation and information extraction capabilities. Previous studies [25][26][27][28] have determined the feasibility of detecting fractures on the basis of neural networks using diagnostic CT images, which can be used for reference in our study of detection of bone condition. As a pioneer work, Armato et al. [27] first used deep learning methodologies to detect vertebral compression fractures in the chest and/or abdomen CT scans. Shortly afterward, Tomita et al. [28] developed a deep neural network framework to detect osteoporotic vertebral fractures on CT scans, which achieved good performance. The aforementioned studies also inspired us to use neural networks to extract features from diagnostic CT images for detection of bone condition.

In this study, we proposed an automatic detection method of bone condition on the basis of CNN for osteoporosis screening. The proposed method can qualitatively provide a judgment on bone condition using diagnostic CT images of the lumbar vertebra. The result of judgment is one of three bone conditions, namely, normal bone mass, low bone mass, and osteoporosis. We performed experiments on clinical data for the detection of bone condition and compared the performance of our method with the performance of radiologists. The details of the proposed method and its evaluation are described in the rest of the paper.

## 2. Materials and methods

In clinical practice, tomography results are inspected by radiologists to provide diagnostic reports for the patients. Radiologists first focus on the objective region of interests (ROIs) and then analyze the ROI features for pathological diagnosis. In our work, similar to the diagnostic procedure of radiologists, we proposed a novel method for automatically detecting bone condition by extracting the features of diagnostic CT images via CNN. On this basis, two functional modules were designed to generate the final diagnostic results of bone condition. In the original diagnostic CT images, the information of bone condition was concentrated in the lumbar vertebra, which was the objective ROI in our study. In this case, the first network module was designed based on a segmentation network to narrow the attention area from the whole CT image to an objective ROI. More importantly, we need to provide the final qualitative diagnostic results of bone condition on the basis of the extracted ROI in diagnostic CT images. Thus, a second network module, which determines the category of bone condition, was designed on the basis of a classification network.

**Figure 1** shows the integrated network including the image segmentation module and image classification module. The image segmentation module uses the original CT image as an input to generate a mark image of the ROI. During the training stage, the segmentation label and the output of the image segmentation module are forwarded to the pixel level loss by the mean absolute error (MAE) [29]. Then, the original CT image and the generated mark image are processed to obtain the segmentation image. The image classification module uses the segmentation image as input to obtain the final classification result, which corresponds to the specific bone condition. During the training stage, the classification label is the bone condition diagnosed by radiologists for original CT image; in addition, the classification label and the output of the image classification module are forwarded by the softmax loss.

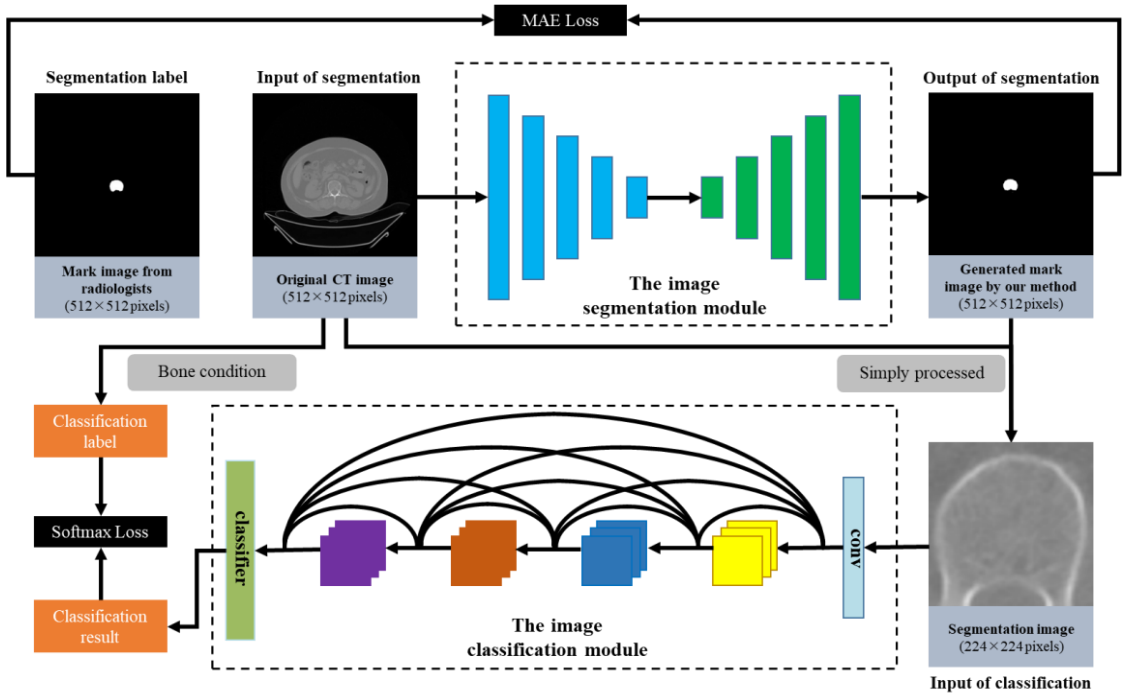

**Figure 1** Schematic of the proposed method. The image segmentation module mainly utilizes the basic structure of U-Net, whereas the image classification module mainly utilizes the basic structure of DenseNet. The blue and green rectangles indicate the encoder and decoder in the segmentation module, respectively. The overlapping rectangular blocks of different colors in the classification module represent extracted image features.

**2.1 Image segmentation module**

In our segmentation task, the dataset of lumbar vertebra CT images with specialized label of lumbar vertebra outlines is very small. However, U-Net has effectively solved the problem of few data because it can train some samples with less data by data augmentation. U-Net consists of a contracting path to capture context and a symmetric expanding path that enables precise localization.

Such a network has been confirmed to perform end-to-end training with less data and achieve good results [20][30]. In our work, we designed a specific network based on U-Net to perform segmentation of the lumbar vertebra in CT images, as shown in **Figure 2**. The network is named as mark-segmentation-Network (MS-Net). The designed MS-Net maintains the encoding-decoding model of U-Net. The features of lumbar vertebra are extracted through the encoding stage, and the features are mapped in the decoding stage to obtain the segmentation results. The original CT images are used as input, and its corresponding mark images are used as labels to train the network model.

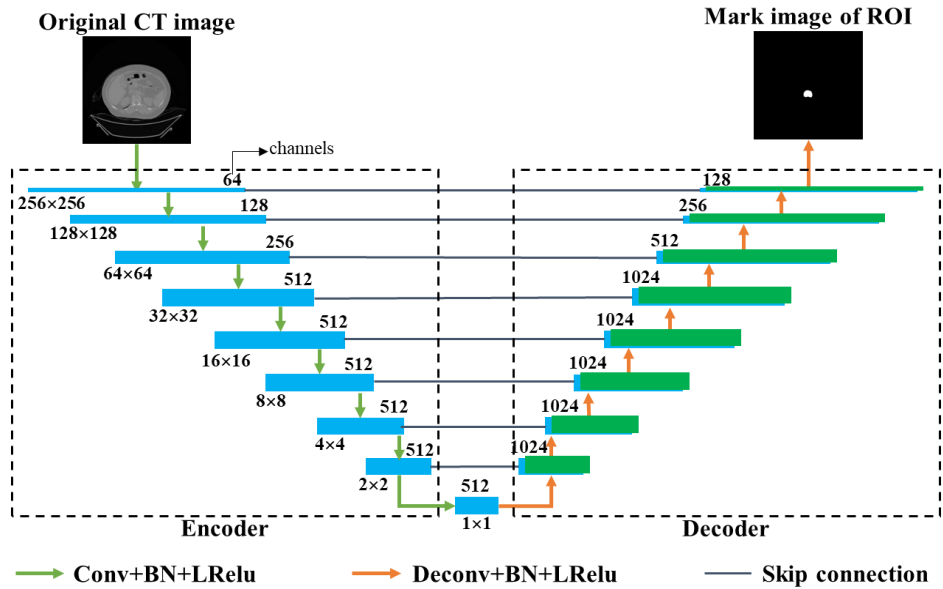

**Figure 2** Designed MS-Unet used for the segmentation module. The blue and green rectangles indicate the image blocks generated by the network layers in MS-Unet.

The encoder of MS-Net extracts image features from the input data by using nine convolutional layers. The first three convolutional layers have 64, 128, and 256 channels with a filter kernel size of 4×4, and the overlapping stride is 2. The fourth to the ninth convolutional layer has 512 channels with a filter kernel size of 4×4, and the overlapping stride is 2. The activation functions of convolution layers are leaky rectified linear unit (LReLU) [31] with a slope of 0.2. The decoder of MS-Net aims to obtain a mark image of lumbar vertebra from the acquired feature information of CT image. The decoder consists of the corresponding eight deconvolutional layers. The first six deconvolutional layers have 512 channels with a filter kernel size of 4×4, and the overlapping stride is 2. The seventh to the last deconvolutional layer has 256, 128, and 64 channels with filter kernel size of 4×4, and the overlapping stride is 2. The skip connection connects the corresponding encoder and decoder layers to help the decoder better obtain the details of the mark image. The original CT images have a uniform size of $i×512×512×1$, where $i$ is the batch size of the training data. The output image size of MS-Net is the same as the input size.

**2.2 Image classification module**

In image classification tasks, when applying CNN to extract high-level semantic features of the image, the network should be deep enough to reduce the resolution of the feature map to acquire abstracted features. Given the sensitivity of private information and the need for precise and professional labeling by radiologists, data with specialized lumbar vertebra outlines and bone condition diagnosis results are scarce. Therefore, the designed CNN should be deep and have relatively fewer parameters. Compared with other CNNs with the same input size, DenseNet has fewer parameters [32], and it weakens the gradient vanish problem of deep networks. Some problems also arise when DenseNet is directly used in the classification task of bone condition. Unlike ImageNet [33], which often contains millions of images with labels, the dataset of lumbar vertebra CT images with bone condition diagnosis results is small. Therefore, DenseNet architecture for ImageNet has more parameters when applied to the classification task of bone condition and is easy to lead to network overfit. Hence, reduction of network training parameters is a requisite. Through decreasing the network parameters and maintaining the connected manner of DenseNet, we employed the Bone-Conditions-Classification-Network (BCC-Net) to discriminate the categories of bone condition, as shown in **Figure 3**.

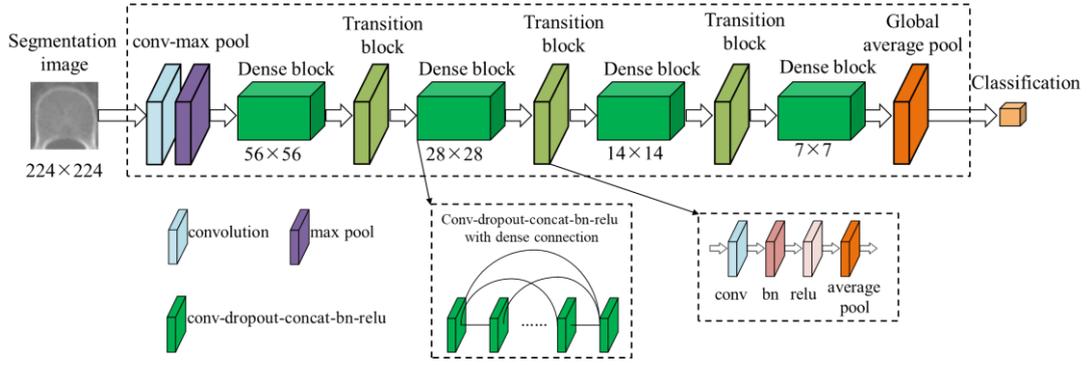

**Figure 3** Designed M-DenseNet used for the classification module.

The mark image is processed together with the CT image to obtain a corresponding segmentation image, which is the input of the classification module. We cropped the ROI area of the original CT image and scaled it to 224×224, which is suitable for the classification module. We selected DenseNet architecture for ImageNet (DenseNet-121) as a modified template. The growth of the training parameters of DenseNet is mainly caused by the number of convolutional layers in the dense block and the growth rate $k$, which refers to the output number of feature maps. The growth rate $k$ is often limited to a small integer to prevent the network from growing too wide and to improve parameter efficiency. To mitigate overfitting, we reduced the number of dense blocks to 1, 2, 4, and 2 in sequence. The growth rate $k$ is set to 12, which is consistent with the setting of CIFAR dataset [34] classification task to reduce the number of training parameters. A 1×1 convolutional layer followed by a 2×2 average pooling layer is introduced as a bottleneck layer to reduce the number of input feature maps because of the concatenation operation, thereby improving computational efficiency. We also limited the output of the bottleneck layer to a smaller integer compared with DenseNet-121 to fit the small dataset of lumbar vertebra CT images. Similar to the original DenseNet, a global average pooling and a softmax classifier are performed at the end of the last dense block. The output of the final classification layer is changed according to the categories of bone condition, which is set to 3 in our research. The detailed network configuration is shown in

**Table 1**.

Table 1 DenseNet architectures for ImageNet and the dataset of lumbar vertebra CT images. Each "conv" layer shown in the table corresponds to the sequence conv-bn [35]-relu [36].

| Layers | Output Size | DenseNet-121 (k=32) | BCC-Net (k=12) |
|---|---|---|---|
| Convolution | 112×112 | 7×7 convolution, stride 2 | |
| Pooling | 56×56 | 3×3 max pooling, stride 2 | |
| Dense Block(1) | 56×56 | $\begin{bmatrix} 1\times 1 \text{ conv} \\ 3\times 3 \text{ conv} \end{bmatrix} \times 6$ | $\begin{bmatrix} 1\times 1 \text{ conv} \\ 3\times 3 \text{ conv} \end{bmatrix} \times 1$ |
| Transition Layer(1) | 56×56 | 1×1 convolution | |
| | 28×28 | 2×2 average pooling, stride 2 | |
| Dense Block(2) | 28×28 | $\begin{bmatrix} 1\times 1 \text{ conv} \\ 3\times 3 \text{ conv} \end{bmatrix} \times 12$ | $\begin{bmatrix} 1\times 1 \text{ conv} \\ 3\times 3 \text{ conv} \end{bmatrix} \times 2$ |
| Transition Layer(2) | 28×28 | 1×1 convolution | |
| | 14×14 | 2×2 average pooling, stride 2 | |
| Dense Block(3) | 14×14 | $\begin{bmatrix} 1\times 1 \text{ conv} \\ 3\times 3 \text{ conv} \end{bmatrix} \times 24$ | $\begin{bmatrix} 1\times 1 \text{ conv} \\ 3\times 3 \text{ conv} \end{bmatrix} \times 4$ |
| Transition Layer(3) | 14×14 | 1×1 convolution | |
| | 7×7 | 2×2 average pooling, stride 2 | |
| Dense Block(4) | 7×7 | $\begin{bmatrix} 1\times 1 \text{ conv} \\ 3\times 3 \text{ conv} \end{bmatrix} \times 16$ | $\begin{bmatrix} 1\times 1 \text{ conv} \\ 3\times 3 \text{ conv} \end{bmatrix} \times 2$ |
| Classification Layer | 1×1 | 7×7 global average pooling | |
| | | 2D fully-connected, softmax | |

## 3. Results

### 3.1 Data acquisition

The data used in our research are routine diagnostic lumbar vertebra CT images collected from the Department of Radiology in Henan Provincial People's Hospital. We collected 229 cases and selected 213 cases for our experiments, which were mainly obtained from three CT imaging devices, i.e., Ingenuity CT, Ingenuity Flex, and Brilliance16. All cases in our collected dataset have the final diagnosis from the radiation department of the said hospital, which is based on DXA. Categories of bone condition are extracted from the final diagnosis results. In addition, all contours of the lumbar vertebra in the CT images are labeled by experienced and specialized radiologists. The data distribution of various bone conditions is basically balanced. The CT images of lumbar vertebra are gray-level digitized images with the size of 512×512 pixels saved as digital imaging and communications in DICOM format. The dataset consists of original CT images and the corresponding mark images and diagnostic results.

Table 2 Routine diagnostic lumbar vertebra CT image dataset used in our research.

| Bone condition | Train dataset (case) | Test dataset (case) |
|---|---|---|
| Normal bone mass | 50 | 21 |
| Low bone mass | 50 | 21 |
| Osteoporosis | 50 | 21 |
| Total | 150 | 63 |

An overview of our collected data is provided in **Table 2**. We divided the dataset into training and test sets with no repetition according to patient. In the task of image segmentation, the input image is the original lumbar vertebra CT image, and the label is the corresponding mark image. The output image size of the segmentation module is the same as the input size, which is 512×512. The gray value of CT images is standardized to 0-255 when sent to MS-Net. MS-Net can train some samples with less data by data augmentation [37]. Therefore, we enlarged the training dataset by random rotation and random translation. The number of images used for training of MS-Net is 1,800 and that used for testing is 756.

In the task of image classification, the rectangle lumbar vertebra regions are extracted when sent to BCC-Net. For data-driven BCC-Net, training data are extremely insufficient. Thus, we utilized data augmentation to enlarge the dataset and prevent network overfitting. All the samples are augmented by scaling, random translation, random rotation, and a combination of these methods. Scaling is set to the proportion of lumbar vertebra region in the rectangle region of network input. To retain more detailed information of the lumbar vertebra region, the scaling coefficient is set to a small value between 0.8 and 1 to prevent lumbar vertebra deformation. Given that the input size of BCC-Net is 224×224, all the training and test data are resized to fit the network input. Ultimately, the number of images used for training of BCC-Net is 21,600, and that used for testing is 3,024.

## 3.2 Training details and parameters

The proposed method is implemented in Caffe deep learning framework [38]. All experiments are conducted on a workstation equipped with a Tesla K20X GPU (5 GB of memory).

MS-Net is trained using Adam optimizer [39], which is used to update the gradient and network parameters. The learning rate is fixed at 0.002 in the first half of the training process and decreases linearly from 0.002 to 0 in the second half of the training process. The training process has 150 epochs. The first half has 100 epochs, and the second half has 50 epochs. The momentum value is set to 0.9, and the batch size is 64.

BCC-Net is trained using stochastic gradient descent, and the batch size is set to 64. The momentum value is set to 0.9. The initial learning rate is set to 0.01, and the learning strategy is used as follows:

$$L = l \times (1 + \gamma \times N)^{-\alpha}, \tag{1}$$

where $l$ is the initial learning rate, $N$ is the number of training rounds, and $L$ is the updated learning rate. The parameter of $\gamma$ is set to 0.006, and the parameter of $\alpha$ is set to 0.98. The weight decay is set to 0.005 for $L_2$ penalty. The dropout layer [40] with a dropout rate of 0.4 after each convolutional layer (except for the first one) is used to prevent overfitting.

## 3.3 Metrics

We used quantitative metrics to evaluate the performance of the proposed method. For the segmentation result of the first step, we chose the dice index (DI), pixel accuracy (PA), and intersection over union (IOU) to test the performance of MS-Net. The definition of DI, PA, and IOU are as follows:

$$DI = \frac{2 \times TP_p}{2 \times TP_p + FP_p + FN_p}, \tag{2}$$

$$PA = \frac{TP_p}{TP_p + FN_p}, \tag{3}$$

$$IOU = \frac{TP_p}{TP_p + FN_p + FP_p}, \tag{4}$$

where $TP_p$ denotes the number of pixels that are correctly segmented into the lumbar vertebra, $FP_p$ denotes the number of pixels that are incorrectly segmented into the lumbar vertebra, and $FN_p$ denotes the number of pixels that are incorrectly segmented into the background.

For the final classification result, we chose the accuracy, precision, and receiver operating characteristic (ROC) curve [41], which are preferred in most medical image classification work. The definition of accuracy, precision, and ROC curve are as follows:

$$accuracy = \frac{TP_s + TN_s}{TP_s + FP_s + TN_s + FN_s}, \tag{5}$$

$$precision = \frac{TP_s}{TP_s + FP_s}, \tag{6}$$

$$sensitivity = \frac{TP_s}{TP_s + FN_s}, \tag{7}$$

$$specificity = \frac{TN_s}{FP_s + TN_s}, \tag{8}$$

where $TP_s$ denotes the true positive samples, $FP_s$ denotes the false positive samples, $TN_s$ denotes the true negative samples, and $FN_s$ denotes the false negative samples. Accuracy means the proportion of true prediction samples in the total classification samples. Precision means the proportion of true positive prediction samples in the total classification positive samples. ROC curve uses sensitivity, which is represented for true positive rate as vertical axis, and specificity, which is represented for false positive rate as horizontal axis. ROC curve balances the rate of missed diagnosis and misdiagnosis and is a comprehensive evaluation metric. A large area under ROC curve (AUC) means that the classifier has good performance.

## 3.4 Segmentation results

We tested the trained image segmentation module with 756 images from the test dataset of **Table 2** by data augmentation. As shown in **Figure 4**, MS-Net has an excellent segmentation precision and obvious advantage on the shape preservation of different lumbar vertebra. It can identify the location of the lumbar vertebra accurately, confirming the effectiveness of MS-Net for our segmentation task.

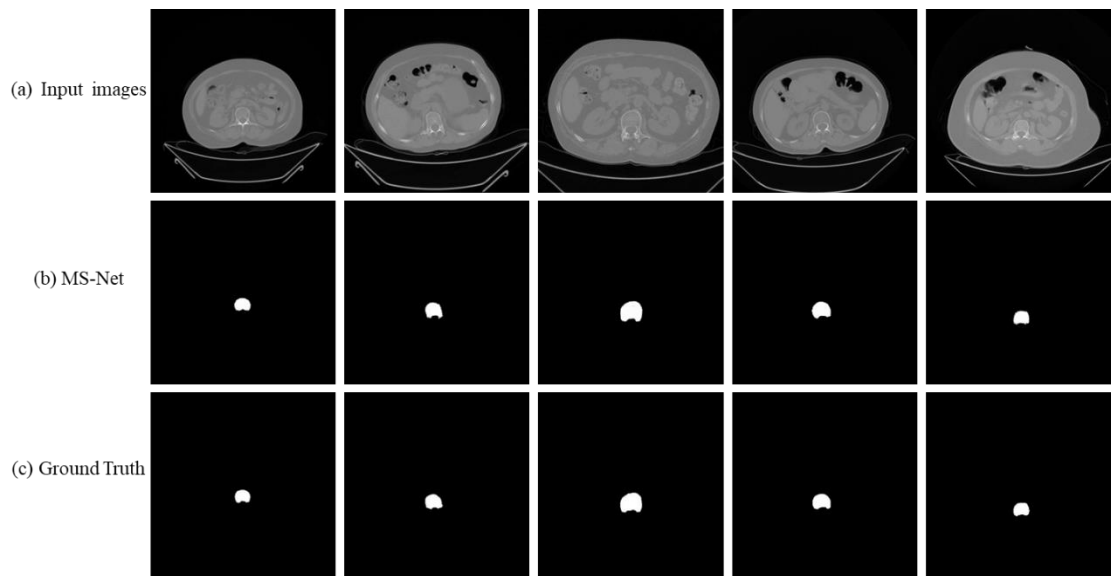

**Figure 4** Example results of image segmentation module.

As shown in **Table 3**, the mean DI of MS-Net on the test dataset is 0.9992, the mean IOU is 0.9984, and the mean PA is 0.9991. The DI, IOU, and PA of MS-Net have an extremely high score. Moreover, we counted the distribution of numerical indicators of segmentation results in the 756 images. As shown in

**Figure 5**, the numerical values of the segmentation results are greater than 0.998. The numerical results indicate that MS-Net can be used to complete the segmentation task.

**Table 3** Quantitative evaluation of segmentation results of the segmentation module.

| Segmentation module | DI | IOU | PA |
| --- | --- | --- | --- |
| MS-Net | 0.9992 | 0.9984 | 0.9991 |

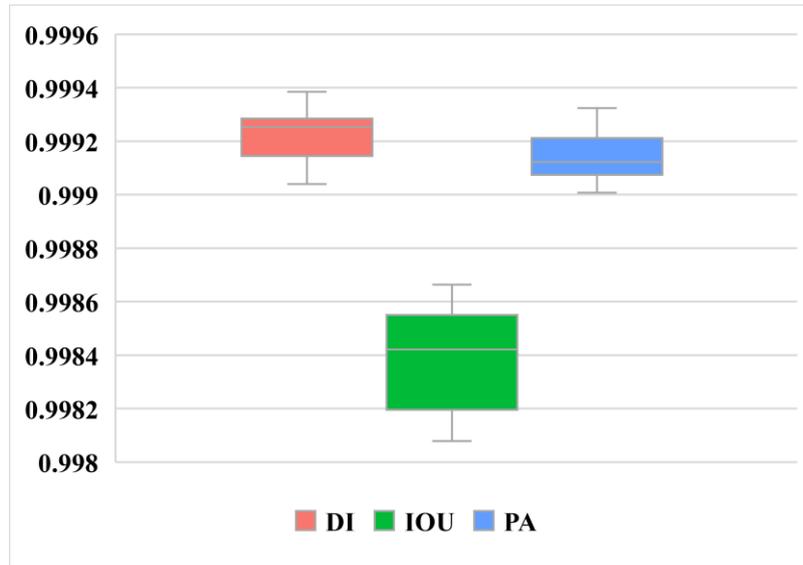

**Figure 5** Distribution of numerical indicators of segmentation results.

**3.5 Classification results**

We used the segmentation module to obtain the corresponding mark images for the test dataset of **Table 2**. Then, the segmentation images were processed together with the original CT images, and 3,024 test images were finally obtained by data augmentation. We tested the trained image classification module with 3,024 images. The precision of the classification and ROC curve of the different bone conditions are shown in **Table 4** and **Figure 6**. The precision of the classification for normal bone mass, low bone mass, and osteoporosis is 80.57%, 66.21%, and 82.55%, respectively. The AUC has similar distribution with precision. From the numerical results, the performance of the model is ordinary for low bone mass but better for normal bone mass and osteoporosis. This situation can also be seen intuitively from the ROC curve.

**Table 4** Classification performance for different bone conditions.

| Bone condition | Precision | AUC |
| --- | --- | --- |
| Normal bone mass | 0.8057 | 0.9715 |
| Low bone mass | 0.6621 | 0.8320 |
| Osteoporosis | 0.8255 | 0.9467 |

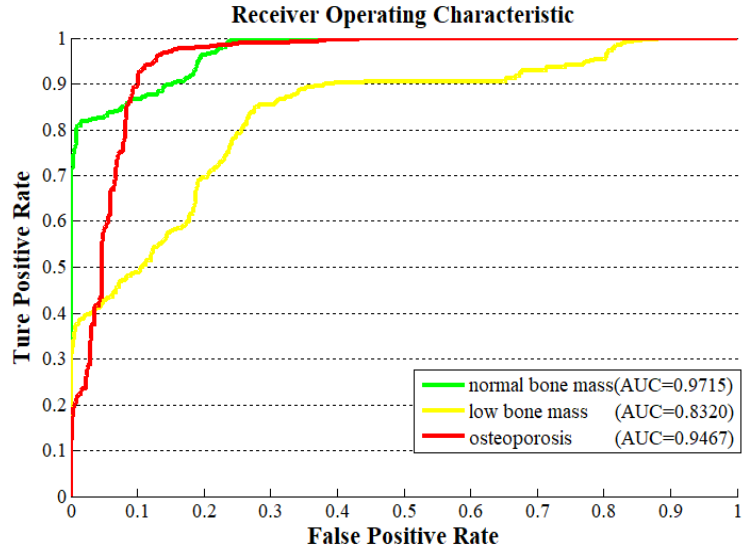

**Figure 6** ROC curve of three different bone conditions.

This situation is in line with objective laws. In our research, the low bone mass is between normal bone mass and osteoporosis; its classification complexity will be higher than that of the other two categories. In practice, we statistically analyzed the number of samples for each category in the classification results. The data distribution is shown in **Figure 7**.

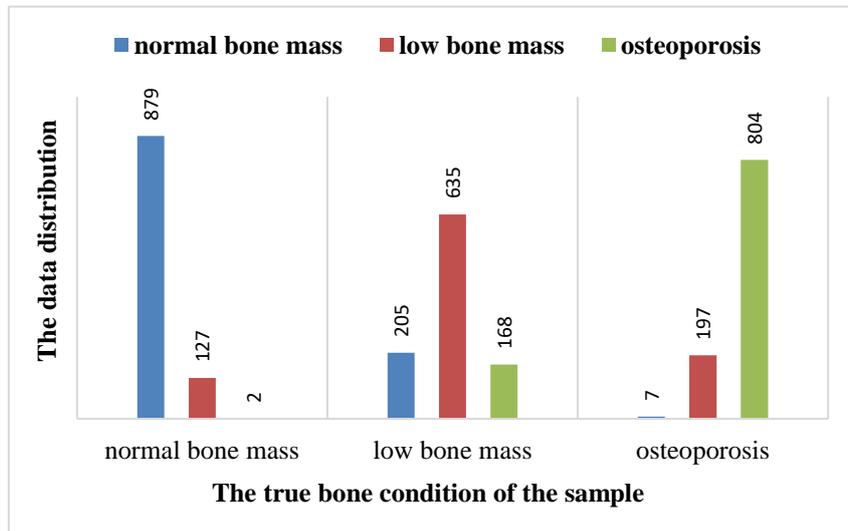

**Figure 7** Distribution of samples in each category in the classification result.

In **Figure 7**, the horizontal axis represents the true bone condition of the sample, and the vertical axis represents the data distribution of the classification results by our method. For example, in 1,008 samples with normal bone mass, the classification result shows that 879 samples are classified correctly and 129 samples are classified incorrectly. Among the samples with incorrect classification, 98.4% are diagnosed as low bone mass, and only 1.6% are diagnosed as osteoporosis. From the data distribution, the classification of normal bone mass samples and osteoporosis samples is mainly interfered by the adjacent category, whereas low bone mass is obviously interfered by the other two categories. Therefore, classification of low bone mass samples is more difficult.

We also compared the proposed BCC-Net with the original DenseNet-121 to validate the necessity of decreasing network parameters. The pre-trained DenseNet-121 model on ImageNet dataset serves as the initial parameter, and the network output number is changed from 1,000 to 3. As shown in **Table 5**, the proposed BCC-Net obviously has a better performance than DenseNet-121 both in accuracy and AUC. For small dataset of lumbar vertebra CT images, the pre-trained model has poor performance due to considerable image differences of the two datasets. This finding indicates the requirement of reducing network training parameters. We also compared the results of different amounts of dense block layers. The two compared counts of dense block layers are 1, 1, 2, 1 and 2, 4, 6, 4 in sequence, named DenseNet-1 and DenseNet-2. As shown in **Table 5**, BCC-Net achieves higher scores in accuracy and AUC. Therefore, the amount of dense block layers in BCC-Net is more suitable for our classification task.

Table 5 Classification performance of Network with different dense block amounts.

| Classification module | Accuracy | AUC |
|---|---|---|
| DenseNet-121 | 0.5498 | 0.6842 |
| DenseNet-1 | 0.6581 | 0.8120 |
| DenseNet-2 | 0.6412 | 0.7923 |
| **BCC-Net** | **0.7665** | **0.9167** |

We also compared the performance between the proposed BCC-Net and practicing radiologists on the test dataset in **Table 2**. The specific diagnostic results of radiologists are shown in Appendix A. Three radiologists from the Department of Radiology in Local People's Hospital analyzed the original CT images to provide the diagnostic results through their comprehensive clinical experience.

To protect the privacy of the radiologists, they are represented as radiologist-1, radiologist-2, and radiologist-3. As shown in **Table 6**, the performance of the proposed method is better than that of the three radiologists. In terms of physical examinations, the proposed method may be more valuable than the empirical diagnosis of radiologists.

Table 6 Comparison of the classification performance between BCC-Net and practicing radiologists.

| | Metrics | Radiologist-1 | Radiologist-2 | Radiologist-3 | BCC-Net |
|---|---|---|---|---|---|
| Precision | Normal bone mass | 0.6250 | 0.6471 | 0.7500 | **0.8057** |
| | Low bone mass | 0.3571 | 0.3333 | 0.3929 | **0.6621** |
| | Osteoporosis | 0.6316 | 0.6250 | 0.6316 | **0.8255** |
| Accuracy | | 0.5079 | 0.4921 | 0.5556 | **0.7665** |

The detection of bone condition has great significance for the prevention of osteoporosis, and it is difficult for radiologists to discriminate the categories of bone condition on the basis of lumbar vertebra CT images. As shown in **Figure 8**, (a) is the sample of normal bone mass, (b) is the sample of low bone mass and (c) is the sample of osteoporosis. The lumbar vertebra CT images have very similar representations, but they belong to different categories. Moreover, some differences in pixel-level image details exist. BCC-Net connects all the layers with each other, so the final classification layer makes the decision on the basis of all the extracted features from the preceding layers, resulting in excellent classification performance.

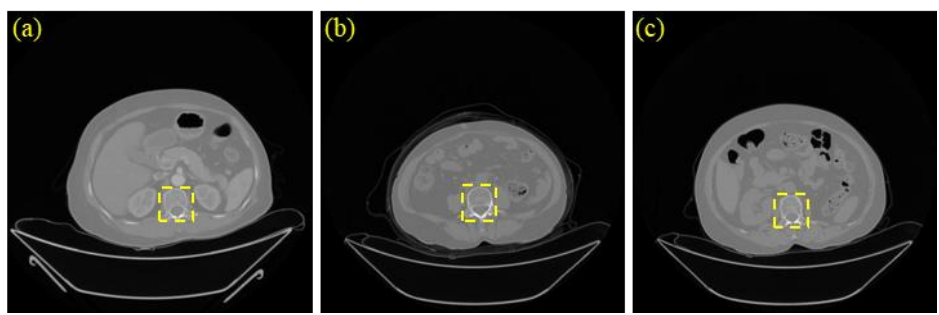

Figure 8 Lumbar vertebra CT images with different bone conditions.

In clinical practice, diagnostic CT images can be obtained from routine physical examinations or diagnosis of diseases. Then, we can determine the probability of which bone condition of physical examinees belongs to each category using our method. On the basis of this probability, we drew a pie chart to help radiologists provide a report of the bone condition for the physical examinees. For example, as shown in the first pie chart in the first row of **Figure 9**, the processing results of a correct classification sample in the classification module are three probability values in our experiment, which are 0.9933, 0.0065, and 0.0002. These three values represent the probability that the sample is normal bone mass, low bone mass, and osteoporosis, respectively. According to this probability distribution, radiologists can judge that the risk of osteoporosis is extremely low for the physical examinees.

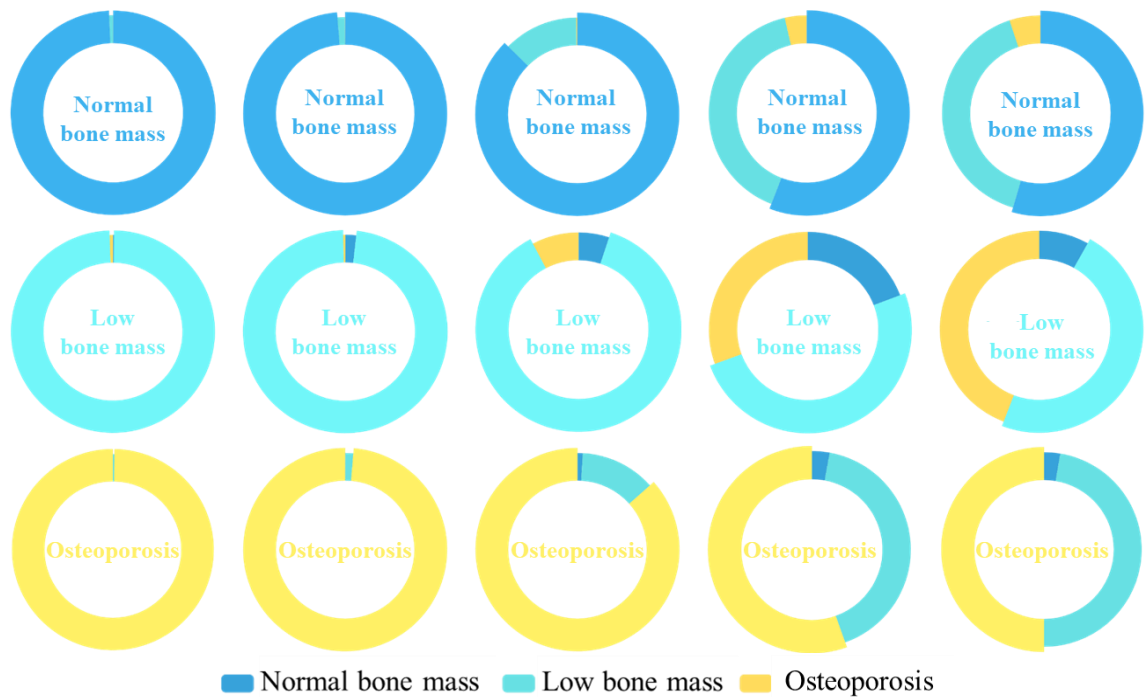

Figure 9 Figure 10 Pie charts showing the classification probability for part of samples. The results displayed in the middle of each pie chart represent the diagnosed bone condition by the proposed method. The true bone condition of the pie chart in the first line is normal bone mass, the second line is low bone mass, and the last line is osteoporosis.

The results of the first, second, and third columns in **Figure 9** all have one category with a large proportion of probability. Although the results in the fourth and fifth columns of the figure are also correctly diagnosed, the category with the highest probability does not have an overwhelming advantage for other categories. Therefore, when the result of the diagnosis is normal bone mass but the proportion of probability is not high, radiologists can judge the risk of osteoporosis for the physical examinees or suggest that the physical examinees need more exercise. When the result of the diagnosis is low bone mass or osteoporosis and the proportion of probability is high, radiologists can offer some treatment plans in the early stage and recommend that the physical examinees regularly measure their BMD. When the result of the diagnosis is low bone mass or osteoporosis and the probability is comparable with the other categories, further examinations can be performed on the physical examinees to confirm their bone condition. Then, the radiologists will provide the corresponding treatment plans.

## 4. Discussion and conclusions

In this study, we proposed a novel two-step method for osteoporosis screening by automatic detection of bone conditions via diagnostic CT images. First, we used MS-Net to effectively identify and segment the lumbar vertebra in the CT images, which could narrow the attention area and avoid extracting meaningless features for classification module. Then, we used BCC-Net to classify the segmentation images, which could obtain the results of the bone condition corresponding to the original CT images. Ultimately, we achieved automatic detection of bone condition with an accuracy of 76.65% and an AUC of 0.9167. The proposed automatic detection method can potentially reduce the time and the manual burden on radiologists for detecting bone condition and reduce the economic burden on physical examinees. By integrating our method into routine CT examinations, we can provide an extra report of the bone condition after obtaining the l umbar vertebra CT image of the physical examinees. This will improve the detection rate of bone condition and help the physical examinees to prevent bone mass loss and early development of osteoporosis, thereby promoting their healthy development.

In this work, we not only provided a qualitative diagnostic result of the bone condition but also presented the probability distribution of physical examinees under three bone conditions in a pie chart. Compared with qualitative diagnostic results, this pie chart more clearly shows the current bone condition of the physical examinee, and the clinical radiologist can offer more targeted preventive measures or treatments using the pie chart. In particular, when the probability of two bone conditions in the pie chart is relatively large and very close, the network will only provide a qualitative diagnosis, which is not reasonable. In this case, the radiologist needs to provide more objective suggestions on the basis of the probability distribution of the pie chart. In addition, the dataset we collected is small in our study, making the loss function more difficult to converge during the training stage. Collecting more data for training may enhance the robustness of the proposed method.

We used the deep learning method to qualitatively detect the bone condition for osteoporosis screening. Although BMD measurement using DXA and QCT can provide quantitative results of bone condition diagnosis, our method can improve the diagnosis of bone condition and help physical examinees to prevent bone mass loss and the early development of osteoporosis. In future studies, some improvements can be done. In our method, we only considered the features extracted from lumbar vertebra CT images, and other clinical characters might have an important influence on the detection of bone condition. In the future, more image features and clinical characters are going to be exploited to determine the bone condition.

# Appendix A

| | The bone condition of samples | The diagnostic results of radiologists | | |
|---|---|---|---|---|
| | | Radiologist-1 | Radiologist-2 | Radiologist-3 |
| 1 | Normal bone mass | Normal bone mass | Normal bone mass | Normal bone mass |
| 2 | Low bone mass | Osteoporosis | Low bone mass | Osteoporosis |
| 3 | Normal bone mass | Low bone mass | Low bone mass | Low bone mass |
| 4 | Osteoporosis | Osteoporosis | Osteoporosis | Osteoporosis |
| 5 | Low bone mass | Low bone mass | Osteoporosis | Low bone mass |
| 6 | Normal bone mass | Low bone mass | Normal bone mass | Normal bone mass |
| 7 | Osteoporosis | Low bone mass | Osteoporosis | Osteoporosis |
| 8 | Normal bone mass | Normal bone mass | Low bone mass | Normal bone mass |
| 9 | Normal bone mass | Low bone mass | Low bone mass | Normal bone mass |
| 10 | Low bone mass | Normal bone mass | Normal bone mass | Low bone mass |
| 11 | Osteoporosis | Osteoporosis | Low bone mass | Low bone mass |
| 12 | Low bone mass | Low bone mass | Low bone mass | Osteoporosis |
| 13 | Osteoporosis | Low bone mass | Low bone mass | Low bone mass |
| 14 | Osteoporosis | Osteoporosis | Osteoporosis | Osteoporosis |
| 15 | Normal bone mass | Normal bone mass | Normal bone mass | Low bone mass |
| 16 | Low bone mass | Low bone mass | Normal bone mass | Normal bone mass |
| 17 | Osteoporosis | Osteoporosis | Low bone mass | Osteoporosis |
| 18 | Low bone mass | Osteoporosis | Low bone mass | Low bone mass |
| 19 | Normal bone mass | Normal bone mass | Normal bone mass | Normal bone mass |
| 20 | Low bone mass | Low bone mass | Osteoporosis | Low bone mass |
| 21 | Osteoporosis | Low bone mass | Osteoporosis | Low bone mass |
| 22 | Osteoporosis | Osteoporosis | Low bone mass | Osteoporosis |
| 23 | Normal bone mass | Low bone mass | Low bone mass | Low bone mass |
| 24 | Low bone mass | Normal bone mass | Normal bone mass | Normal bone mass |
| 25 | Normal bone mass | Normal bone mass | Low bone mass | Normal bone mass |
| 26 | Normal bone mass | Low bone mass | Normal bone mass | Low bone mass |
| 27 | Low bone mass | Low bone mass | Osteoporosis | Osteoporosis |
| 28 | Normal bone mass | Osteoporosis | Low bone mass | Low bone mass |
| 29 | Osteoporosis | Osteoporosis | Osteoporosis | Osteoporosis |
| 30 | Low bone mass | Osteoporosis | Low bone mass | Low bone mass |
| 31 | Osteoporosis | Low bone mass | Low bone mass | Low bone mass |
| 32 | Normal bone mass | Low bone mass | Low bone mass | Osteoporosis |
| 33 | Low bone mass | Normal bone mass | Low bone mass | Normal bone mass |
| 34 | Osteoporosis | Osteoporosis | Osteoporosis | Osteoporosis |
| 35 | Osteoporosis | Low bone mass | Low bone mass | Osteoporosis |
| 36 | Normal bone mass | Low bone mass | Normal bone mass | Low bone mass |
| 37 | Normal bone mass | Normal bone mass | Normal bone mass | Normal bone mass |
| 38 | Osteoporosis | Osteoporosis | Low bone mass | Low bone mass |
| 39 | Low bone mass | Low bone mass | Normal bone mass | Low bone mass |
| 40 | Normal bone mass | Normal bone mass | Low bone mass | Normal bone mass |
| 41 | Osteoporosis | Low bone mass | Osteoporosis | Low bone mass |
| 42 | Osteoporosis | Osteoporosis | Low bone mass | Osteoporosis |
| 43 | Low bone mass | Low bone mass | Normal bone mass | Low bone mass |
| 44 | Low bone mass | Osteoporosis | Low bone mass | Osteoporosis |
| 45 | Normal bone mass | Low bone mass | Osteoporosis | Low bone mass |
| 46 | Osteoporosis | Osteoporosis | Osteoporosis | Osteoporosis |
| 47 | Low bone mass | Normal bone mass | Low bone mass | Low bone mass |
| 48 | Normal bone mass | Normal bone mass | Normal bone mass | Normal bone mass |
| 49 | Osteoporosis | Normal bone mass | Low bone mass | Low bone mass |
| 50 | Osteoporosis | Osteoporosis | Low bone mass | Osteoporosis |
| 51 | Low bone mass | Low bone mass | Osteoporosis | Osteoporosis |
| 52 | Normal bone mass | Low bone mass | Low bone mass | Low bone mass |
| 53 | Low bone mass | Osteoporosis | Low bone mass | Osteoporosis |
| 54 | Osteoporosis | Low bone mass | Osteoporosis | Low bone mass |
| 55 | Low bone mass | Low bone mass | Low bone mass | Normal bone mass |
| 56 | Normal bone mass | Normal bone mass | Normal bone mass | Normal bone mass |
| 57 | Normal bone mass | Low bone mass | Normal bone mass | Normal bone mass |
| 58 | Low bone mass | Osteoporosis | Low bone mass | Low bone mass |
| 59 | Low bone mass | Low bone mass | Osteoporosis | Low bone mass |
| 60 | Osteoporosis | Osteoporosis | Osteoporosis | Low bone mass |
| 61 | Normal bone mass | Normal bone mass | Normal bone mass | Normal bone mass |
| 62 | Osteoporosis | Low bone mass | Low bone mass | Osteoporosis |
| 63 | Low bone mass | Normal bone mass | Normal bone mass | Low bone mass |